# Valley Degeneracies in (111) Silicon Quantum Wells

*Neerav Kharche[1*], Seongmin Kim[1], Timothy B. Boykin[3], Gerhard Klimeck[1,2]*

[1]*Birck Nanotechnology Center, Network for Computational Nanotechnology,
Purdue University, West Lafayette, Indiana 47907-1285, USA*

[2]*Jet Propulsion Laboratory, California Institute of Technology, Pasadena, CA 91109*

[3]*Department of Electrical and Computer Engineering, University of Alabama at Huntsville, Huntsville, AL, 35899,*
* *Corresponding authors: email nkharche@purdue.edu*

(111) Silicon quantum wells have been studied extensively, yet no convincing explanation exists for the experimentally observed breaking of 6 fold valley degeneracy into 2 and 4 fold degeneracies. Here, systematic $sp^3d^5s^*$ tight-binding and effective mass calculations are presented to show that a typical miscut modulates the energy levels which leads to breaking of 6 fold valley degeneracy into 2 lower and 4 raised valleys. An effective mass based valley-projection model is used to determine the directions of valley-minima in tight-binding calculations of large supercells. Tight-binding calculations are in better agreement with experiments compared to effective mass calculations.

Silicon nanostructures exhibit a plethora of interesting physical phenomena due to the 6 fold valley degeneracy of the bulk conduction band. Silicon devices are being pursued for spin based quantum computing[1] and spintronics[2] due to their scaling potential and integratability within the industrial nanoelectronic infrastructure. Relative energies and degeneracies of spin and valley states are critical for device operation in these novel computing architectures[1,2], and conventional metal−oxide−semiconductor field−effect transistors (MOSFETs) that often involve the formation of a 2 dimensional electron gas (2DEG) at the semiconductor-insulator interface. Valley degeneracy of the 2DEG is highly dependent on the interface orientation. (100) Si quantum wells (QWs) show lower 2 fold and raised 4 fold valley degeneracy while (110) Si QWs show lower 4 fold and raised 2 fold valley degeneracy. The origin of these valley degeneracies is well understood and the experimental observations are in agreement with the effective mass based theoretical predictions[3].

Valley degeneracy in (111) Si QWs should be 6 according to standard effective mass theory. Experimental measurements on (111) Si/SiO$_2$ MOSFETs, however, show a conflicting valley degeneracy of 2 and 4[3-5]. Recently 2-4 valley-splitting has also been observed in magneto-transport measurements performed on hydrogen terminated (111) Si/vacuum field effect transistors[6]. Previously proposed theory of local strain domains[4] can not explain this splitting since the Si-vacuum interface is stress free. The splitting is also unlikely to be a many-body phenomenon[7].

Careful imaging of the surface morphology shows the presence of mono-atomic steps (miscut) on the (111) Si surface[6,8] as well as at the Si/SiO$_2$ interface in (111) Si MOSFETs[8,9]. Atomistic models such as tight-binding are needed to accurately model the electronic structure of miscut QWs. Through systematic tight-binding calculations of flat and miscut (111) Si QWs we show that the surface miscut leads to the 2-4 degeneracy breaking and resolve the conflict between theory and experimental observations. To reduce the computational burden associated with searching the whole Brillouin-zone for valley minima, an effective mass based valley-projection model[10] tailored to miscut (111) surfaces is used. Electronic structure calculations are performed using the general purpose NEMO-3D Code[11].

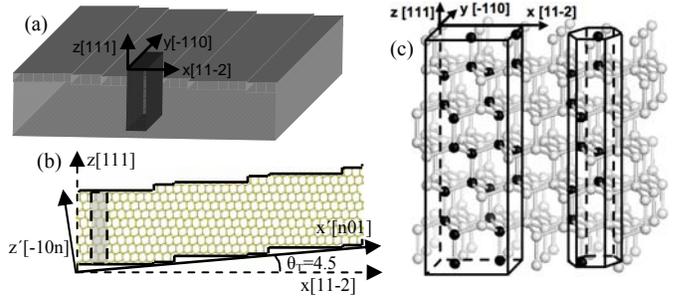

Fig. 1: (a) Schematic of a miscut (111) quantum well **(QW)**. A rectangular unit cell is repeated in space to build the miscut quantum well. (b) Atomistic view of a unit cell of a 4.5° miscut (111) Si QW. The reduced symmetry along the direction perpendicular to steps results in a larger unit cell. The smallest repeated **miscut QW unit** cell along $[11\bar{2}]$ direction has 6 steps. (c) Atomistic view of a flat (111) Si **QW**. The rectangular unit cell is larger than the hexagonal primitive unit cell.

Hydrogen terminated (111) Si surfaces as well as the (111) Si/SiO$_2$ interfaces have mono-atomic steps (Fig 1(a,b)). For simplicity, the steps are assumed to run perpendicular to the $[11\bar{2}]$ direction. This surface morphology can be implemented by repeating the rectangular unit cell of Fig 1(c) in the miscut $[11\bar{2}]$ direction. The in-plane unit cell dimensions are $a_x = \sqrt{3/2}\,a_{Si}$ and $a_y = \sqrt{1/2}\,a_{Si}$, where $a_{Si}$ is Si lattice constant[12]. The advantage of using a rectangular unit cell is two-fold. First, the rectangular geometry simplifies the underlying mathematics and implementation of the periodic boundary conditions in the bandstructure calculation. Second, the surface miscut can be easily implemented by applying shifted boundary condition to the rectangular supercell.

The Brillouin Zone and 6 degenerate valleys in the bandstructure of the hexagonal primitive unit cell of a flat (111) Si QW (Fig. 1(c)) are shown in Fig 2(a). The 2 bulk valleys along $[001]$ direction are projected along $[11\bar{2}]$ direction while the remaining 4 bulk valleys along $[100]$ and $[010]$ are symmetrically projected in each of the four quadrants [10]. The 2 valleys along $[11\bar{2}]$ are labeled as $A$ and remaining 4 valleys are labeled as $B$. The rectangular unit cell is larger than the hexagonal primitive unit cell (Fig 1(c)). The Brillouin Zone of the rectangular unit cell of Fig. 1(b) is given by $\{(k_x,k_y): -\sqrt{2/3}\,\pi/a_{Si} \leq k_x \leq \sqrt{2/3}\,\pi/a_{Si},\; -\sqrt{2}\,\pi/a_{Si} \leq k_y \leq \sqrt{2}\,\pi/a_{Si}\}$. It is smaller than the hexagonal Brillouin Zone in Fig. 1(a) and 2 $A$ type valleys are folded.

The miscut surface morphology can be conveniently implemented by extending the rectangular unit cell in one direction (Fig. 1(b,c)). The unit cell shown in Fig. 1(b) has a miscut angle of 4.5°. In the $(x,y,z)$ co-ordinate system of Fig. 1(a) the surface normal of this QW is along $[\bar{1}0n]$ direction. This direction is related to the miscut angle by $\theta_T = \tan^{-1}(a_z/na_x)$, where $a_x = \sqrt{3/2}\,a_{Si}$ and $a_z = \sqrt{3}\,a_{Si}$. Typical miscuts range from 0.1-8°. Before going to the experimental 0.2° miscut[6] we illustrate the essential physics and reduce the computational burden significantly by studying the effect the miscut of 13°, which can be investigated in a system extending six unit cells along the miscut direction and has a smaller supercell compared to 0.2° miscut. These unit cells are schematically shown along with energy contours of their lowest conduction bands in Fig. 2(c,d). Only the positive quadrant is shown. The flat QW supercell of Fig 2(c) is 6 times longer in the x-direction compared to the rectangular unit cell used to plot the bandstructure in Fig. 2 (b) resulting in zone-folding of this supercell to 1/6th. In the first Brillouin zone $A$-type valleys lie along the $k_x$-direction and the $B$ type valleys lie along the solid zig-zag line. In the repeated-zone scheme the bandstructures along the solid zig-zag line is the same as that along the dotted straight line. The bands corresponding to $A$ and $B$ type valleys clearly show that in a flat (111) Si QW both valleys are degenerate while in a miscut QW two $A$-type valleys have lower energy compared to four $B$-type valleys (Fig. 3(d)). This degeneracy breaking is the effect of different confinement effective masses of $A$- and $B$-type valleys in a miscut QW.

Experimentally relevant QWs have typical miscuts ranging from 0.1-8°. As the miscut angle becomes smaller the size of the unit cell increases. For example, the unit cell of 23 nm thick 13° miscut QW contains 2,100 atoms while a 0.2° miscut QW requires 119,100 atoms. An effective mass theory can be used to determine the directions of valley minima thereby reducing the computational burden associated with searching the whole Brillouin zone. Here we outline the valley projection model[10] as applied to determine the directions of valley minima of miscut (111) Si QWs.

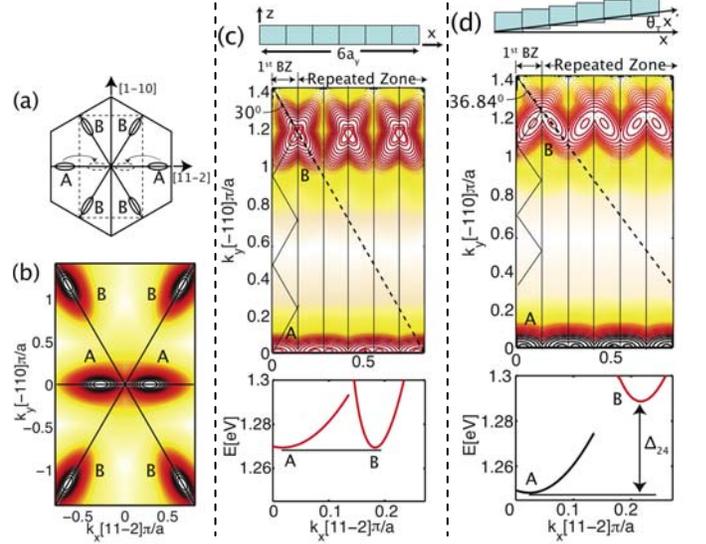

Fig. 2: (a) Brillouin zone of the primitive unit cell of a **flat** (111) Si **quantum well** (QW). 6 degenerate valleys are shown. (b) Brillouin zone and bandstructure of a (111) Si QW plotted using the rectangular unit cell of Fig. 1 (b). The bandstructure is folded as shown schematically in (a). (c) Bandstructure of a flat (111) Si **QW** in repeated zone scheme plotted using a supercell which contains 6 small cells in x-direction. The bandstructure of Fig. (b) is folded in the 1st Brillouin zone. A type valleys along $k_x$-direction and B type valleys along the dotted line are degenerate. (d) Bandstructure of a 13° miscut (111) Si **QW** in repeated zone scheme. A-type valleys along $k_x$-direction are lower in energy than B-type valleys along the dotted line. **I**n the flat **QW** the dotted line subtends an angle of 30° with the negative y-axis while this angle is 36.84° in a miscut **QW**.

Consider the miscut QW unit cell of Fig. 1(c), the rotation matrix $R$ from bulk valley co-ordinate system to $(x',y,z')$ co-ordinate system is given by

$$R = \frac{1}{\sqrt{n^2+1}} \begin{bmatrix} n & 0 & 1 \\ 0 & 1 & 0 \\ -1 & 0 & n \end{bmatrix} \begin{bmatrix} 1/\sqrt{6} & 1/\sqrt{6} & -\sqrt{2/3} \\ -1/\sqrt{2} & 1/\sqrt{2} & 0 \\ 1/\sqrt{3} & 1/\sqrt{3} & 1/\sqrt{3} \end{bmatrix} \quad (1)$$

In the effective mass formalism the sub-bands of this QW are given by [10]

$$E_i(k_{x'},k_y) = E_i^0 + \frac{1}{2}\hbar^2 \left[ \left( w_{x'x'} - \frac{w_{x'z'}^2}{w_{z'z'}} \right)k_{x'}^2 + 2\left( w_{x'y} - \frac{w_{x'z'}w_{yz'}}{w_{z'z'}} \right)k_{x'}k_y + \left( w_{yy} - \frac{w_{yz'}^2}{w_{z'z'}} \right)k_y^2 \right] \quad (2)$$

where, the reciprocal effective mass matrix $[W]$ is given by

$$[W]_{ij} = \sum_\alpha R_{i\alpha} R_{j\alpha} [M_0]^{-1}_{\alpha\alpha}, \quad i,j,\alpha \in \{x',y,z'\} \quad (3)$$

here $[M_0]$ is the effective mass matrix in the bulk valley co-ordinate system. The position of the sub-band minimum, $E_i^0$, is determined by the confinement effective mass $m_{z'} = 1/w_{z'z'}$ and the confinement potential in the direction perpendicular to the QW surface. In this formalism two A-type valleys which lie along [100] direction are projected along $k_{x'}$ while the remaining four B-type valleys are projected along the directions which subtend angles $\pm\varphi$ with $\pm k_y$ axis. The angle $\varphi$ can be determined by rotating the co-ordinate system $(k_{x'}, k_y)$ such that the cross term in eq. (2) vanishes. This angle is given by $\varphi = \frac{1}{2}\tan^{-1}\left(\frac{c}{a-b}\right)$ where $a$, $b$ and $c$ denote coefficients of terms $k_{x'}^2$, $k_y^2$ and $k_{x'}k_y$ respectively. One of these four directions which lie in the positive quadrant is shown for flat and miscut QWs in Fig. 2.

While in a flat (111) QW confinement effective masses, $m_{z'}$, are the same for A- and B-type valleys, a miscut alters these masses such that $m_A > m_B$ resulting in a broken degeneracy of lower 2 fold and raised 4 fold. Although the effective mass theory can explain the origin of $\Delta_{2-4}$ splitting more sophisticated methods such as tight-binding are needed to accurately model the effect of mono-atomic surface steps on the electronic structure. The $\Delta_{2-4}$ splitting increases with the miscut angle due to increasing difference between confinement effective masses of A- and B-type valleys. Both effective mass and tight-binding models show this trend, the effective mass model, however, gives smaller splitting compared to the tight-binding model. As shown in Fig. 3 the step morphology of the QW surface modulates the wavefunction which in turn influence energy levels to give rise to $\Delta_{2-4}$ splitting.

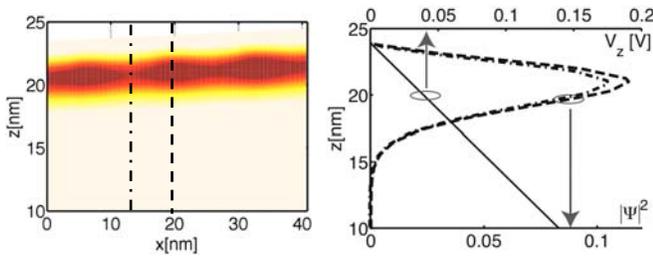

Fig. 3: (a) A typical wavefunction of a 4.5° miscut (111) Si QW of Fig. 1(b) at the valley minimum. Thickness of the QW is 23 nm, however, only the near surface portion having appreciable wavefunction magnitude is shown. (b) Electrostatic potential and wavefunction cut along two dotted lines in Fig (a). Confining potential due to constant electric field (10 MV/m) pulls the wavefunction to the surface. Surface steps modulate the wavefunction amplitude and modify energy levels which give rise to the $\Delta_{2-4}$ splitting.

A 0.2° miscut resembles closely to that of the experiment[6]. The unit cell of a QW of this miscut is $L_x$=264.07 nm and $L_y$=0.38 nm long in x- and y- directions, respectively. A constant z-directed electric field of 10 MV/m which corresponds to the electron density of $6.5\times10^{11}$ cm$^{-2}$ is assumed[6]. To avoid any truncation effects of electronic domain on eigenvalues a QW thickness of $L_z$=23 nm is simulated. This is the smallest unit cell which can be repeated in xy-plane to generate a 23 nm thick 0.2° miscut QW. This unit cell contains around 0.1 millon atoms which makes it computationally expensive to search the whole 2 dimensional Brillouin zone for valley minima. To reduce the computational burden, the valley-projection method described above is used to identify the directions of valley minima. The valley minima of A-type valleys occur along the $k_x$-direction while valley minima of B-type valleys occur at an angle $\varphi=\pm30.13°$ to $k_y$ axis. The confinement effective masses, $m_z$, for A- and B-type valleys are $0.2608m_0$ and $0.2593m_0$ (where $m_0$ is the mass of the free electron) respectively. The $\Delta_{2-4}$ valley-splitting in a 0.2° miscut QW calculated using an effective mass and the sp$^3$d$^5$s$^*$ tight-binding models are 1.25 (108 µeV) and 3.98 K (343 µeV) respectively. An analytical formula for energy levels in triangular potential wells is used to estimate the splitting in the effective mass model[10]. The splitting reported from the temperature dependence of the longitudinal resistance[6] is 7K (604 µeV) which shows that the tight-binding calculation matches closer to experiments compared to the effective mass calculation.

In conclusion, the miscut morphology of the (111) Si surface is shown to be the origin of breaking of 6 fold valley degeneracy into lower 2 and raised 4 fold valley degeneracies. Atomistic basis representation such as tight-binding is needed to capture the effect of wavefunction modulation at mono-atomic steps on the electronic structure. Compared to effective mass the tight-binding calculations are found to match closer to experimentally measured $\Delta_{2-4}$ splitting. Additional surface phenomena could be responsible for enhanced $\Delta_{2-4}$ splitting reported in experiments[6].

This work was supported by Semiconductor Research Corporation and the Army Research Office. The work described in this publication was carried out in part at the Jet Propulsion Laboratory, California Institute of Technology under a contract with the National Aeronautics and Space Administration, and Jet Propulsion Laboratory. nanoHUB.org computational resources operated by the Network for Computational Nanotechnology, funded by the National Science Foundation were used.